\documentclass[twocolumn]{aa}
\usepackage{natbib}
\usepackage{color}
\usepackage{ragged2e}
\usepackage[pdfpagelabels=false]{hyperref}	
\hypersetup{colorlinks=true,linkcolor=blue,citecolor=blue,filecolor=blue,urlcolor=blue,}
\usepackage[varg]{txfonts}
\usepackage{graphicx,rotating}
\usepackage[normalem]{ulem}
\usepackage{colortbl}
\usepackage{xcolor}
\usepackage{bbold}

\bibpunct{(}{)}{;}{a}{}{,} 

\definecolor{darkolivegreen}{rgb}{0.33, 0.42, 0.18}
\definecolor{salmon}{rgb}{0.95,0.5,0.25}

\begin{document}

\title{In-situ globular clusters in alternative dark matter Milky Way galaxies: a first approach to fuzzy and core-like dark matter theories}

\titlerunning{Globular clusters in alternative dark matter Milky Way galaxies}
\author{Pierre Boldrini \inst{1} and Paola Di Matteo \inst{1}}

\offprints{Pierre Boldrini, \email{pierre.boldrini@obspm.fr}}
\institute{$^{1}$ LIRA, Observatoire de Paris, Université PSL, Sorbonne Université, Université Paris Cité, CY Cergy Paris Université, CNRS, 92190 Meudon, France}
\authorrunning{Boldrini et al.}
\date{accepted by A$\&$A}

\abstract{We present a first analysis of the dynamics of in-situ globular clusters (GCs) in Milky Way (MW)-like galaxies embedded in fuzzy dark matter (FDM) halos, combining cosmological assembly histories from the TNG50 simulation with dedicated orbital integrations and analytical models. GC populations are initialized with identical distributions in normalized $E$-$L_{z}$ in matched CDM and FDM halos. In a universe dominated by FDM, we identify three distinct regimes for the in-situ GC population depending on the particle mass $m_{22} \equiv m_{\chi}/ 10^{-22}~\mathrm{eV}$. For $m_{22} < 7$, baryons dominate the inner potential, which remains steep and centrally concentrated, confining GC orbits to a narrow region and producing less massive, more compact systems than in CDM. For $m_{22} \sim 7$, GC properties resemble those in CDM, with similar mass and spatial distributions. For $m_{22} > 7$, the dark matter becomes both compact and globally dominant, generating a deeper and more extended gravitational potential that supports a wider range of stable GC orbits, resulting in more massive and spatially extended GC systems. Finally, we extend our framework to make predictions for GC populations in alternative DM models, including warm dark matter and self-interacting dark matter, in both MW-like and dwarf galaxies. Our findings demonstrate that in-situ GC systems offer a sensitive and independent probe of the underlying DM physics, opening new avenues for observational constraints with upcoming Euclid.}

\keywords{Dark matter - galaxy dynamics - globular clusters - Milky Way - methods: orbital integrations - cosmological simulations}
\maketitle



\section{Introduction}

Although its existence is firmly established through its gravitational effects on baryonic matter, the nature of dark matter (DM) remains unknown. The standard cosmological model, cold DM (CDM), continues to face challenges on galactic scales (see \cite{2017ARA&A..55..343B,2021Galax..10....5B} for reviews), thereby motivating alternative models such as warm DM (WDM) \citep{1980PhRvL..45.1980B,1994PhRvL..72...17D,2000PhRvD..62f3511H,2006PhRvD..73f3506A,2005PhRvD..71f3534V}, self-interacting DM (SIDM) \citep{2000PhRvL..84.3760S,1992ApJ...398...43C,1995ApJ...452..495D}, fuzzy DM~\citep{2000NewA....5..103G,2000PhRvL..85.1158H}, or even primordial black holes (see \cite{2025arXiv250215279C} for a review). In this context, globular clusters (GCs) are particularly valuable as probes of galactic dynamics. These compact stellar systems are essentially DM-free, relatively low in mass, and roughly an order of magnitude more numerous than satellite galaxies (171 Milky Way (MW) GCs versus 11 classical satellites). In contrast to satellite galaxies, GCs also span a wide range of galactocentric distances, from 0.7 to 200 kpc in the MW. These ancient relics encode key information about the assembly history of the MW and are strongly influenced by the gravitational effects \citep{2022MNRAS.513.3925R,2022arXiv220411861R} and intrinsic properties of DM by extension.

European space missions such as Gaia~\citep{Gaia21}, and more recently Euclid \citep{2025A&A...697A...1E}, has revolutionized our ability to study GCs. Gaia has provided full six-dimensional phase-space information for nearly all Galactic GCs~\citep{Vasiliev21}. Euclid, whose first observations have recently been released~\citep{2024arXiv241217672U,2025A&A...697A...8M,2025A&A...697A..13K,Sai2025V1,Sai2025V2}, is expected to transform our understanding of GC populations in nearby galaxies, with the anticipated detection of up to half a million extragalactic GCs out to 100 Mpc~\citep{2021sf2a.conf..447L,Voggel25}. Thanks to its high spatial resolution, Euclid will characterize an unprecedented number of GCs around galaxies of all types, in both high- and low-density environments, including MW-like galaxies in the local universe. Early observations in the Fornax cluster have already recovered over 95$\%$ of the known GCs, and resolved GCs as compact as $r_h = 2.5$ pc at 20 Mpc ~\citep{Sai2025V2}. This large dataset offers a unique opportunity to place new constraints on the nature of DM.

In this study, we focus on the FDM model, proposed as a viable alternative to CDM (see \cite{2025arXiv250700705E} for a review). DM is then assumed to be an ultra-light scalar field with no self-interaction in the non-relativistic limit, referred to as FDM. This scalar field is assumed to be made up of very light particles with a mass of $m_{\chi}=0.1-30 \times 10^{-22}$ eV. Galactic studies, such as the analysis of the thickening of cold stellar streams \citep{Amorisco18}, suggest $m_{22} > 1.5$, while studies of the MW disk indicate a boson mass around $m_{22} \sim 0.5 - 0.7$ \citep{2023MNRAS.518.4045C}. These results are in tension with some of the more stringent constraints derived from the Lyman-$\alpha$ forest, which limit the mass to the range $m_{22} =7 - 20$ \citep{2017PhRvL.119c1302I,2017PhRvD..96l3514K,2019MNRAS.482.3227N,2017MNRAS.471.4606A,2021PhRvL.126g1302R}. However, the latter rely on sensitive assumptions regarding the intergalactic medium, and their validity remains debated, especially due to the neglect of effects such as radiative transfer \citep{Hui17,BO19, 2019MNRAS.489.3456G, 2020PhRvD.101l3026S,2023MNRAS.518.4045C,2024PhRvD.110d3534P}. Such ultra-light DM particles have a characteristic wavelength called the de Broglie wavelength: 
\begin{equation}
    \lambda_{B}= 1.19 \; m_{22}^{-1} \;\left(\frac{100\,\mathrm{km/s}}{v_{rel}}\right) \mathrm{kpc},
    \label{eqz1}
\end{equation}
where $v_{\mathrm{rel}}$ is the characteristic velocity of infalling objects and $m_{22} = m_{\chi} / 10^{-22}\, \mathrm{eV}$. Equation~\eqref{eqz1} shows that wavelengths of a few kiloparsecs are of astrophysical scale, where quantum effects become significant. As a result, they can alter the classical dynamical properties predicted by CDM \citep{2000PhRvL..85.1158H,2009ApJ...697..850W,2007JCAP...06..025B}. In essence,
\begin{equation}
\text{if} \quad m_{22} \searrow \quad \Rightarrow \quad \lambda_{B} \nearrow,
\end{equation}
meaning that quantum effects extend over increasingly larger regions of the galaxy as the particle mass decreases.

Integrating FDM into cosmological simulations remains a major challenge due to its intrinsic wave-like nature \citep{Schive14,Schive16,Mocz19,Veltmaat20,May21,2020arXiv200704119M,2022MNRAS.511..943C,May23,2023MNRAS.522.1451N}. FDM simulations are significantly more computationally demanding than their CDM counterparts, primarily because they require solving the Schrödinger–Poisson equations that govern FDM dynamics. One of the key limitations arises from the time step $\Delta t$, which scales quadratically with spatial resolution. As a result, increasing resolution leads to prohibitively small time steps. Moreover, the spatial resolution must be fine enough to resolve the de Broglie wavelength (see Equation~\eqref{eqz1}), regardless of the particle velocity. Failing to meet this condition results in substantial numerical errors, such as an incorrect halo mass function. These stringent numerical requirements restrict full FDM simulations to relatively small cosmological volumes (e.g., 1.4~Mpc$/h$ at $z=0$ in \cite{Schive14}) and often to high redshifts only (e.g., $z=3$ for a 10~Mpc$/h$ box in \cite{May21,May23}), making robust statistical studies difficult. For instance, a simulation with a grid size of $8192^3$ cells evolved down to $z=0$ would require at least $10^7$ CPU hours and would still be limited to box sizes smaller than 10~Mpc$/h$ for the relevant mass range~\citep{May21}.

Despite the growing interest in FDM, studies explicitly coupling its dynamics with that of baryonic systems such as galactic disks, GCs, or stellar streams~\citep{2018arXiv180800464A,Chiang23} remain scarce, largely due to the numerical challenges discussed above. These limitations hinder the development of a statistically robust understanding of GC dynamical evolution within a fully consistent FDM framework. To overcome these difficulties, we adopt a post-processing approach applied to cosmological simulations, inspired by the methodology developed by \cite{Boldrini25}. This long-term strategy aims to build statistically meaningful GC models that can fully leverage the upcoming observational potential of the Euclid mission. Our method combines existing cosmological simulations with orbital integration techniques to study the hierarchical assembly of GC populations. We use tagging techniques, where GCs are "tagged" at high redshift within galaxies based on physically motivated prescriptions. Using this approach, \cite{Boldrini25} follow the trajectories of approximately 18,000 GCs across a sample of 198 MW–like galaxies extracted from the hydrodynamical cosmological simulation TNG50~\citep{Nelson19a,Nelson19b,Pillepich19}, tracing their evolution from formation to $z = 0$. This post-processing method is particularly advantageous due to its low computational cost, enabling the study of large statistical samples of galaxies.

In this paper, we investigate the impact of FDM on the dynamics of in-situ GCs in MW-like galaxies, from redshift $z = 2$ to $z = 0$. Our approach combines orbital integration techniques and analytical models with realistic cosmological assembly histories extracted from the TNG50 simulation, enabling us to track the long-term evolution of GC populations in time-dependent galactic potentials. We focus exclusively on the in-situ GC population, those clusters formed within the main progenitor of the host galaxy, as tracers of FDM-induced dynamical effects. In particular, we explore how GC survival and orbital structure depend on the FDM particle mass $m_{22}$, which governs the shape and depth of the central potential. The paper is organized as follows. Section 2 reviews the relevant properties of FDM at galactic scales. Section 3 describes the construction of time-evolving FDM galactic potentials based on TNG50 mass accretion histories. In Section 4, we detail the initial conditions for the GC populations and the orbital integration scheme used to evolve them forward in time. Section 5 presents our first results on the evolution of GC systems across different DM scenarios. Section 6 examines the dependence of GC survival, and spatial and mass distribution on $m_{22}$, identifying three distinct regimes for the in-situ GC population at $z = 0$. In Section 7, we generalize our approach to alternative DM models, including warm dark matter (WDM) and self-interacting dark matter (SIDM), and explore their implications for GC populations in both MW-like and dwarf galaxies. Finally, Section 8 summarizes our main conclusions and discusses the prospects for testing these predictions with current surveys such as Euclid.

\section{FDM physics at galaxy scales}

To simulate the dynamics of GCs in galaxies composed of FDM, it is essential to account for the specific properties of FDM that distinguish it from the standard CDM model \citep{Marsh19, Hui17, Lancaster20, Amorisco18, Chiang23, Schive14, Schive16}.  
In summary, the dynamics of GCs in an FDM context may be affected by:
\begin{itemize}
    \item a different DM halo density profile, with a central core instead of a cusp,
    \item a fluctuating potential that induces dynamical heating of infalling objects,
    \item a substantial modification of dynamical friction due to the wave-like nature of FDM via its intrinsic density fluctuations (granules),
    \item a significantly reduced population of low-mass subhalos.
\end{itemize}
These factors directly impact the local gravitational potential, orbital energy dissipation, and tidal forces, all of which are critical for modeling the long-term evolution and survival of GCs.

\subsection{FDM distribution}

\begin{figure}
    \centering
    \includegraphics[width=\linewidth]{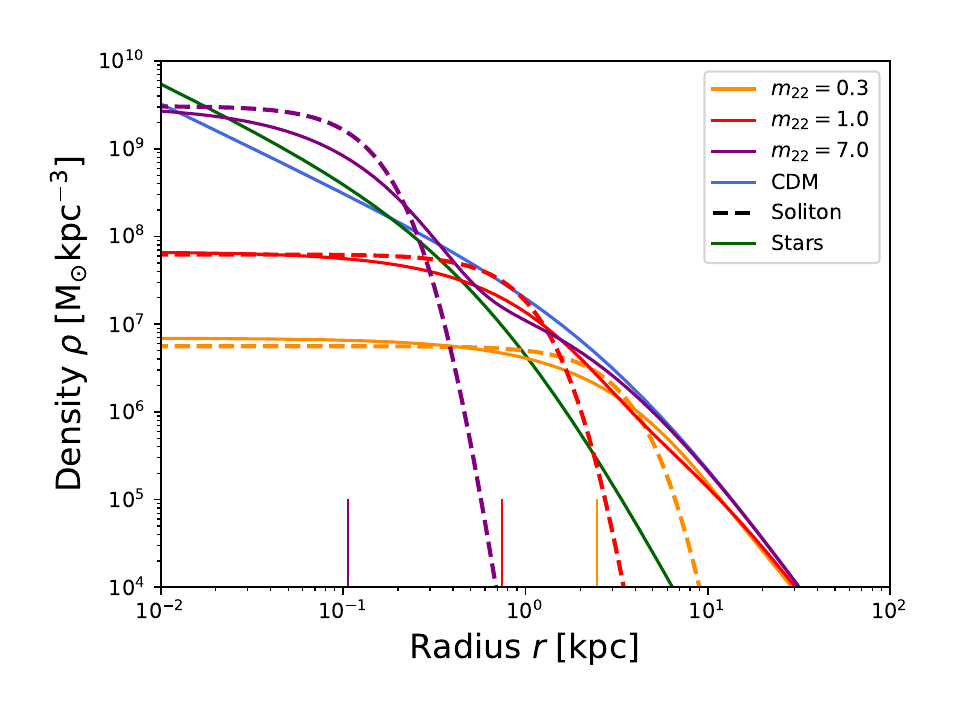}
    \caption{DM profiles of a $10^{10}~M_{\odot}$ halo at $z=0$ for different values of $m_{22}$ as in \cite{Spergel20}. The dashed lines show the central solitonic profiles. The blue line shows the NFW profile of a $10^{10}~M_{\odot}$ halo at $z=0$. The solid lines show our model for the full halo profile, which is a combination of the FDM profile transitioning to an NFW profile, described by Equations~\eqref{galpy1} and ~\eqref{galpy2}. The vertical lines indicate the size of the FDM core radius. The green solid line corresponds to the typical stellar density profile for this halo mass, extracted from the TNG50.}
    \label{fig0}
\end{figure}

Unlike the predictions of CDM, which generally features central density cusps in halos, where the density sharply increases toward the center and is often modeled by an NFW profile \citep{NFW}, FDM halos are characterized by a dense, nearly constant-density central core known as a soliton. It represents a stable, lowest-energy solution of the Schrödinger–Poisson equation, which describes the behavior of FDM under the combined influence of gravity and quantum pressure. Cosmological simulations of FDM have shown that the density profile of the innermost central region of DM halos at redshift $z$ follows the form~\citep{Schive14}:
\begin{equation}
    \rho(r) = \frac{\rho_0}{\left(1 + 0.091\,(r/r_{\mathrm{c}})^2 \right)^8},
\end{equation}
where the central density is given by:
\begin{equation}
\rho_0(z, r_c, m_{22}) = \frac{0.019 \times 10^9}{(1+z)^3} m_{22}^{-2} r_c^{-4} \; M_{\odot}\,\mathrm{kpc}^{-3},
\label{rhoFDM}
\end{equation}
and the core radius is defined as:
\begin{equation}
r_c(z, M_h, m_{22}, \Omega_m) = \frac{1.6}{\sqrt{1 + z}} \left( \frac{\zeta(z, \Omega_m)}{\zeta(0, \Omega_{m0})} \right)^{-1/6} \left( \frac{M_h}{10^9} \right)^{-1/3} m_{22}^{-1},
\label{rcFDM}
\end{equation}
with the redshift-dependent virial overdensity:
\begin{equation}
\zeta(z, \Omega_m) = \frac{18\pi^2 + 82(\Omega_m - 1) - 39(\Omega_m - 1)^2}{\Omega_m}.
\end{equation}
The presence of this central core is a key feature of FDM in galaxies. The properties of the core, in particular its size ($r_c$) and central density ($\rho_0$), strongly depend on the FDM particle mass ($m_{22}$) and the host halo mass ($M_h$). Ultimately, this exotic DM scenario imposes the following relation:
\begin{equation}
\text{if} \quad M_h \searrow \quad \text{or} \quad m_{22} \searrow \quad \Rightarrow \quad r_c \nearrow.
\end{equation}
However, when the de Broglie wavelength $\lambda_B$ is negligible compared to the galactocentric distance under consideration, DM particles can be treated as a classical system. Indeed, at larger radii ($r \gg \lambda_B$), quantum coherence is broken, and the outer regions of FDM halos behave similarly to CDM, with a density profile well approximated by an NFW profile~\citep{Schive14}. Thus, the full density profile of an FDM halo can be expressed as:
\begin{equation}
\rho(r) = \Theta(r_t - r)\, \rho_c + \Theta(r - r_t)\, \rho_{\mathrm{NFW}},
\label{eq1}
\end{equation}
where $\Theta$ is the Heaviside step function, and $r_t$ is the transition radius marking the boundary between the central solitonic core and the outer NFW-like profile.

\subsection{FDM dynamical friction}\label{S22}

By perturbing the matter distribution within a galaxy, an orbiting object generates a gravitational overdensity or wake on its trajectory. The gravitational attraction from this wake exerts a back-reaction on the object, gradually slowing it down. The dynamical friction experienced by a GC, satellite galaxy or a black hole moving through an FDM background differs from the classical \cite{Chandra43} formula, which applies to CDM. This difference arises from the wave-like nature of FDM particles. Quantum interferences of FDM waves within the halo produces ubiquitous, stochastic density fluctuations, often referred to as granules. Constructive and destructive interference creates regions of higher (granules) and lower density. The typical size of these granules is on the order of the de Broglie wavelength, $\lambda_B$. Therefore, if the characteristic size of the system $l \ll \lambda_B$, dynamical friction is modified due to oscillations in the gravitational wake density. Overall, dynamical friction in FDM results in a reduction of the force compared to classical Chandrasekhar estimates. Conversely, when $l \gg \lambda_B$, the classical dynamical friction regime is recovered. A convenient way to decide which regime applies in an FDM universe is to consider the quantum Mach number~\citep{Lancaster20}:
\begin{equation}
    \mathcal{M}_Q = 44.56 \left( \frac{v_{\mathrm{rel}}}{1\, \mathrm{km\, s}^{-1}} \right) 
\left( \frac{M_{\mathrm{sat}}}{10^5 M_\odot} \right)^{-1} m_{22}^{-1},
\label{MQN}
\end{equation}
where $M_{\mathrm{sat}}$ is the mass of the infalling object. The regime of interest is $\mathcal{M}_Q \gg 1$, while the classical dynamical friction description corresponds to $\mathcal{M}_Q \ll 1$~\citep{Hui17,Lancaster20,BO19}.

\subsection{DM halo mass cut}

The FDM model also predicts a suppression of small-scale structure formation compared to CDM. This is attributed to a cut-off in the initial power spectrum, which arises from an effective pressure induced by the wave nature of FDM. This pressure counteracts gravity below the Jeans length, preventing the growth of density fluctuations on smaller wavelengths. Consequently, there exists a minimum halo mass below which FDM halos or subhalos either do not form or are significantly less abundant \citep{Marsh14,Mocz17,Chiang21,2021ApJ...916...27D}:
\begin{equation}
M_{\mathrm{halo, min}} = 4.4 \times 10^{7} (1 + z)^{3/4} \left( \frac{\zeta(z, \Omega_m)}{\zeta(0, \Omega_{m0})} \right)^{1/4} m_{22}^{-3/2} M_\odot.
\end{equation}
Thus, the abundance of low-mass subhalos is significantly reduced in FDM compared to CDM. However, it is important to note that the precise characterization of the FDM subhalo mass function at the low-mass end still presents significant uncertainties and potentially conflicting results between different modeling approaches or numerical simulations \citep{Schive16,May23,2022MNRAS.511..943C}. Nevertheless, the overall suppression of low-mass subhalos is a well-established feature of the FDM model.

\subsection{Heating by FDM}

Dynamical heating caused by fluctuations in the gravitational potential is the last distinctive features of FDM \citep{Hui17,BO19,EZ20,2021ApJ...916...27D}. In this alternative DM model, these fluctuations mainly arise from a turbulent density field within FDM halos, characterized by the presence of density granules. Unlike the stochastic and rare perturbations caused by subhalos in CDM, FDM fluctuations are continuous and recurrent, leading to a diffusive heating process that can eventually disrupt GCs over time:
\begin{equation}
\begin{split}
t_{\mathrm{dis}} =\ & \frac{840~\mathrm{Gyr}}{f_{\mathrm{relax}}}
\left( \frac{v_{\mathrm{GC}}}{200~\mathrm{km}~\mathrm{s}^{-1}} \right)^2 
\left( \frac{M_{\mathrm{GC}}}{3 \times 10^5~M_\odot} \right)\\
&\times \left( \frac{0.01~M_\odot~\mathrm{pc}^{-3}}{\rho} \right)^2
\left( \frac{30~\mathrm{pc}}{a} \right)^3 m_{22},
\end{split}
\label{tdis}
\end{equation}
where $\rho$ is the FDM halo density, and $v_{\mathrm{GC}}$ and $M_{\mathrm{GC}}$ are the GC velocity and mass respectively. a denotes the GC core radius. For MW GCs, the median core radius is typically $a=3$ pc \citep{2018MNRAS.478.1520B}. For instance, this FDM-induced heating can cause the thickening of cold stellar streams from dissolving GCs in the MW \citep{Amorisco18}, or it may explain the observed thickening of the Galactic disk over a Hubble time \citep{Chiang23}.

\begin{figure}
    \centering
    \includegraphics[width=\linewidth]{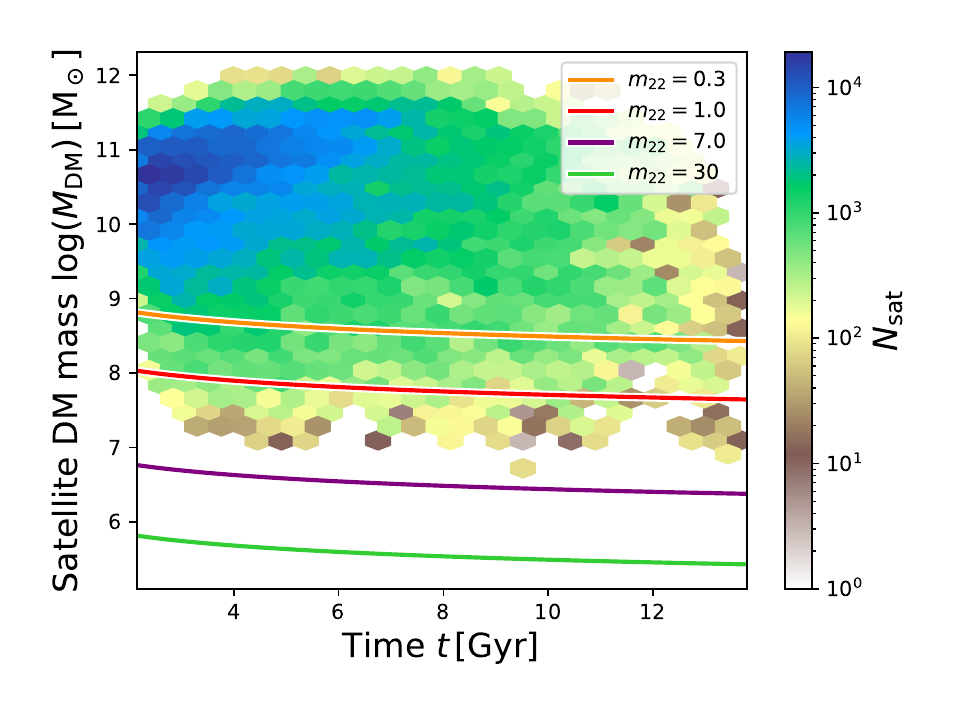}
    \caption{FDM halo mass cut for TNG50 merged satellites: DM mass as a function of time for all merged satellites in the entire TNG50 sample. The hexagonal bins represent at least one satellite. The solid lines represent the low halo mass cutoff for different values of $m_{22}$. The cutoff for $m_{22} = 0.3$ (1.0) removes 6.4\% (1.12\%) of all satellites between $z=2$ and $z=0$. For $m_{22} \geq 7$, no halos are removed.}
    \label{fig01}
\end{figure}

\section{FDM Milky Way-like galaxies}

In \cite{Boldrini25}, we constructed time-evolving gravitational potentials that take into account both the evolution of the MW and its environment, particularly through the accretion of satellite galaxies. More specifically, \cite{Pillepich24} provide the masses and half-mass radii at various redshifts for 198 MW analogues in CDM, as well as for their associated satellites. These satellites were modeled using a \cite{Hernquist90} profile for the stellar component and an NFW profile for the DM \citep{Navarro97}, with parameters calibrated based on properties extracted from TNG50. These MW-like potentials evolve over time, with the mass evolution and characteristic radii integrated into our model. Our current goal is to transform these TNG50 MW analogues from CDM into similar galaxies modeled in the FDM framework, mainly by modifying the density profile while assuming a mass growth comparable to that observed in CDM. In the following, we justify the relevance of this approach. The focus of our study on the in situ populations of GCs within MW-type galaxies allows for certain simplifications in modeling the effects of FDM, as previously presented.

\subsection{FDM Milky Way halo potential over time}

To model an FDM halo, we combined a central solitonic core with an outer power-law envelope. The final FDM profile is thus constructed as the sum of two components using the \texttt{TwoPowerSphericalPotential} profile available in \texttt{galpy}: a solitonic profile approximated by a power law:
\begin{equation}
\rho(r) = \frac{\rho_{01}}{4\pi r_{c1}^3}\left[ 1 + \left( \frac{r}{r_{c1}} \right) \right]^{-\beta},
\label{galpy1}
\end{equation}
and an outer pseudo-NFW envelope, ensuring a physical transition and matching the CDM halo outside the core region, approximated by:
\begin{equation}
\rho(r) = \frac{\rho_{02}}{4\pi r_{c2}^3}\left[ 1 + \left( \frac{r}{r_{c2}} \right) \right]^{-3}.
\label{galpy2}
\end{equation}
We optimized the parameters $(\rho_{01}, \rho_{02}, r_{c1}, r_{c2})$ by minimizing the quadratic error to reproduce the combined soliton + NFW envelope density expected from equation~\eqref{eq1}. Our approximate FDM profiles are shown in Figure~\ref{fig0} for halos of $10^{10}~M_{\odot}$ as in \cite{Spergel20}. Indeed, our profile combination ensures a transition between the core and the NFW envelope with a slope of $-1$ as predicted by FDM. Ultimately, our FDM potential is approximated by a sum of existing \texttt{galpy} potentials, which offers a good compromise between accuracy and performance since they are already implemented in C and thus faster than a newly implemented potential not yet optimized for \texttt{galpy}.

\begin{figure}
    \centering
    \includegraphics[width=\linewidth]{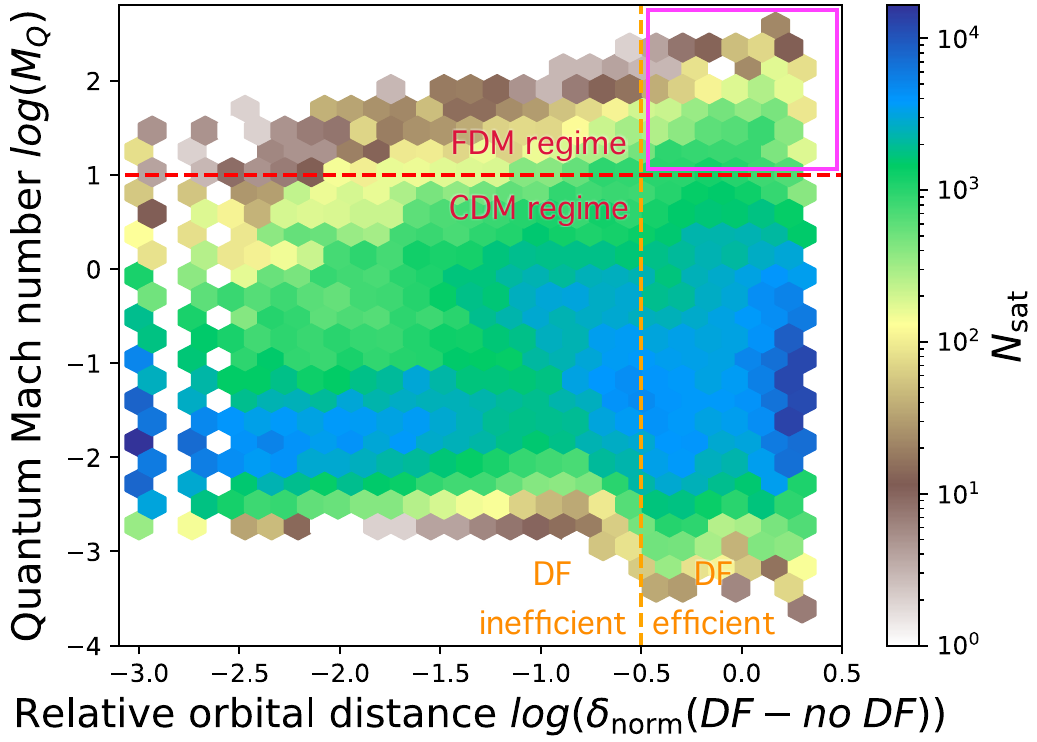}
    \caption{Dynamical friction of merged satellites: Quantum Mach number as a function of the relative orbital distance for all merged satellites in the entire TNG50 sample between $z=2$ and $z=0$ for $m_{22} = 0.3$. The hexagonal bins represent regions containing at least one satellite. The red dashed line shows the adopted boundary between the FDM and CDM regimes, based on the quantum Mach number defined in Equation~\eqref{MQN}. The orange dashed line indicates the threshold where DM dynamical friction becomes efficient, as defined in Equation~\eqref{ODF}. The magenta box highlights the region of interest where FDM dynamical friction must be taken into account, but this only concerns 1.8$\%$ of the satellites.}
    \label{fig02}
\end{figure}

\subsection{TNG50 Milky Way mass growth}

In our study, we took into account the mass growth of the MWs via the merging of satellites predicted by TNG50, but the suppression of low-mass halos predicted by FDM could affect this mass growth. Figure~\ref{fig01} shows that the halo mass cutoff depends on the value of $m_{22}$, but remains negligible for the overall population of merged satellites. Indeed, the cutoff for $m_{22} = 0.3$ (1.0) removes 6.4\% (1.12\%) of the total satellites of the 198 MWs between $z=2$ and $z=0$. Furthermore, for $m_{22} \geq 7$, no halos are removed (see Figure~\ref{fig01}).

Moreover, dynamical friction is also expected to differ in FDM compared to CDM, which could affect the infall of merged satellites within the MWs and modify the mass growth. To quantify this, we computed the quantum Mach number (see Section~\ref{S22}) for the merged satellites of the entire TNG50 MW sample between $z=2$ and $z=0$, which informs us whether dynamical friction is in the FDM or classical regime. In parallel, we measured the relative difference between two trajectories of the same satellite integrated with and without dynamical friction to assess the importance of the latter. It is defined at each time by:
\begin{equation}
   \delta_{\rm norm}(DF - no \;DF)(t) = \frac{2 \, \left| r_{DF}(t) - r_{no\;DF}(t) \right|}{r_{DF}(t) + r_{no\;DF}(t)},
\label{ODF}
\end{equation}
where $r_{DF}$ and $r_{no \; DF}$ are the orbital distances of the satellite from the galactic center with and without dynamical friction in the galactic potential, respectively. Figure~\ref{fig02}, combining this relative orbital distance and the quantum Mach number (see Equation~\eqref{MQN}), shows that only 1.8\% for $m_{22} = 0.3$ of all merged MW satellites within TNG50 between $z=2$ and $z=0$ are in a critical region (magenta square) where FDM dynamical friction is effective and must be taken into account. This percentage decreases as $m_{22}$ increases; for example, we find 0.45\% for $m_{22} = 1.0$. This allows us to assume that the mass growth of the MWs will not be affected by neglecting FDM dynamical friction.

Assuming the TNG50 mass growth for each MW, we can construct an evolving potential between $z=2$ and $z=0$ for the DM halo using redshift-dependent Equations~\eqref{rhoFDM} and ~\eqref{rcFDM} by injecting them into the profile described by ~\eqref{galpy1}. However, this relation may present significant uncertainties and scatter depending on the halo concentration and cosmological parameters. As in \cite{Boldrini25}, we modeled the potentials of the MW-like galaxies by using a \cite{Hernquist90} profile for the stellar component with parameters computed based on properties found in TNG50. This hypothesis is naturally adopted because of the lack of data on the stellar distribution within this exotic DM framework. Detailed studies investigating the interplay between stars and FDM remain limited \citep{Mocz19, Veltmaat20}. Owing to the computational complexity of these numerical simulations, these efforts have been limited to a small number of galaxies, within relatively small volumes (approximately 1.7 to 2.5 Mpc/h) and restricted redshift ranges (up to a maximum of $z=4$), considering only a single FDM particle mass. Nevertheless, since the stellar distribution in the CDM paradigm generally aligns with observations, similar results can be expected for FDM at least at low redshift although the underlying baryonic physics differs. We update the potential 75 times during the orbit integrations. For comparison, we retained the CDM version of the MW potentials used in \cite{Boldrini25}.

\section{In-situ globular clusters in FDM Milky Ways}

\begin{figure*}
    \centering
    \includegraphics[width=\linewidth]{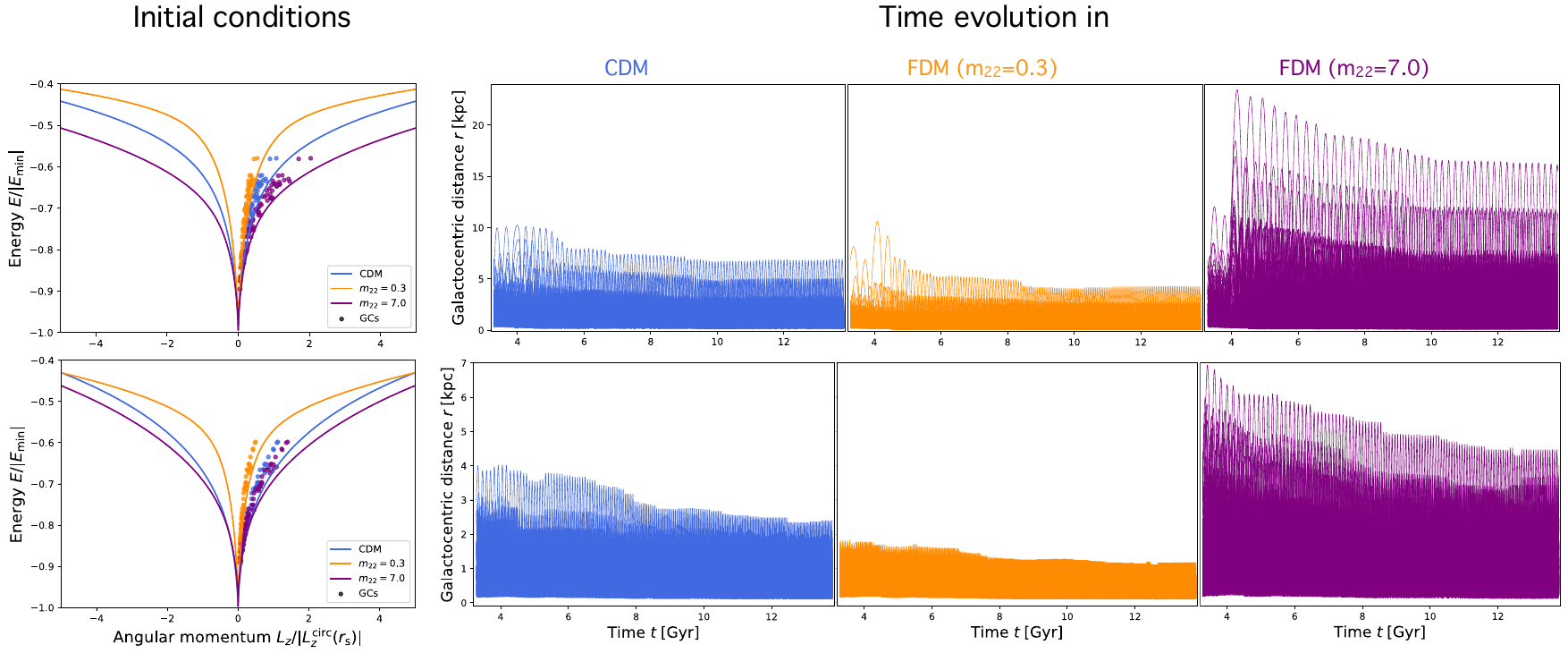}
    \caption{Confined-orbit versus expanded-orbit GCs: 
Initial conditions in left panels: Normalized total energy as a function of the normalized $z$-component of angular momentum at $z=2$ for in-situ GCs (points), used as initial conditions in two different MW host galaxies (IDs 529365 and 572328), under various DM models. The energy is normalized to the absolute value of the minimum of the gravitational potential, and the angular momentum is normalized to the absolute value of that of a circular orbit at the NFW scale radius $r_s$. The colored curves represent the orbital limits allowed by the gravitational potentials in the CDM (blue), FDM with $m_{22}=0.3$ (orange), and FDM with $m_{22}=7.0$ (purple) models. Time evolution in right panels: Galactocentric distances of GCs as function of time from $z=2$ to $z=0$, shown for the three DM models for two MW-like galaxies.} 
    \label{fig05}
\end{figure*}

Here we describe how the in-situ GC system was assigned to each MW progenitor at $z = 2$. The initial number of GCs in each galaxy is defined using the relation from \cite{Burkert20}, which correlates $N_{\mathrm{GC}}$ with the virial mass of the halo:
\begin{equation}
\langle \log N_{\mathrm{GC}} \rangle = -9.58 \pm 1.58 + (0.99 \pm 0.13) \log \left( \frac{M_{\mathrm{vir}}}{M_{\odot}} \right).
\end{equation}
We include uncertainties as random noise in our modeling. We note that applying this relation calibrated on nearby galaxies to systems at $z = 2$ introduces some level of uncertainty. This limitation clearly motivates future zoom-in simulations to study GC formation in high-redshift galaxies, as done by \citet{Sameie2019} in the CDM framework. Indeed, FDM could perturb the compressive shocks of gas and cloud-cloud collisions in the multiphase interstellar medium, thus affecting cluster formation compared to CDM. We derive the initial positions and velocities of in situ GCs from stellar distribution functions based on our sample of progenitor galaxies. As in \cite{Boldrini25}, we use the publicly available code \texttt{AGAMA}\footnote{Available at \url{https://github.com/GalacticDynamics-Oxford/Agama}} to construct equilibrium galaxy models composed of a stellar bulge and a DM halo for our $z=2$ MW progenitor \citep{AGAMA}. Once equilibrium is reached, we sample the corresponding phase-space distribution to generate the positions and velocities of stellar particles, which serve as proxies for GCs in our analysis. We also impose that the $N_{\mathrm{GC}}$ clusters are initially located within the stellar half-mass radius of each galaxy and have angular momentum $L_z > 0.6 L_{\mathrm{circ}}$, where $L_{\mathrm{circ}}$ is the angular momentum of a circular orbit with the same energy. While GCs are placed on disk-like orbits, our Galactic potentials do not include a disk component. In this work, our MWs are only modelled with DM and stellar spherical components. Although these components have been studied in TNG50, they were not catalogued in the same way as the other components for our MW sample to be incorporated into our study \citep{Sotillo22}. This simplification may affect the dynamics of in-situ GCs, for example, by underestimating the mid-plane restoring force.

Regarding the initial conditions of GCs in FDM MWs, we choose to distribute them equivalently in terms of relative gravitational binding energy $E/\left|E_{\rm min}\right|$ and relative angular momentum $L_z/L_z^{\rm circ}(E)$ across the different DM models, as shown in Figure~\ref{fig05}. In other words, the clusters were created with equivalent orbital parameters in each potential. This disc-like initial configuration for the in-situ population ensures a fair comparison between gravitational potentials that have very different structures (cores versus cusps). This approach guarantees that the GCs probe dynamically similar regions in each model, allowing us to isolate the impact of the DM nature on their orbital evolution and mass loss. In our modelling, we do not assign a mass or size to the GCs. Instead, we treat them as test particles, since dynamical friction is negligible over our integration timescale ($\sim$ 11 Gyr) due to the large mass ratio between the enclosed MW mass and typical GC masses. This allows us to omit the specific implementation of FDM dynamical friction, which will be necessary for modeling the ex-situ population. For the mass loss calculation, each GC is initially assigned a mass of $10^6~M_{\odot}$. The mass decay applied in post-processing does not impact GC orbits, since dynamical friction is negligible. We also neglect dynamical heating of GCs induced by density fluctuations in FDM galaxies. Indeed, Figure~\ref{fig04} shows that this disruption process is exceedingly slow for MW GCs, even in the central region, with timescales ranging from 30 to 400 Gyr within the inner 1 kpc, assuming $m_{22} = 0.3$ for an FDM halo with a total mass of $10^{12}~M_{\odot}$. We computed this disruption time using Equation~\eqref{tdis} for the 171 GCs in the MW as provided by Gaia \citep{Vasiliev21}, assuming a relaxation efficiency factor $f_{\rm relax}$ that must be less than or equal to 1 \citep{Hui17}. In addition, the timescale of the process increases with the boson mass $m_{22}$. It is important to note that, for field stars in the MW, this heating will be particularly effective, given that their masses are 3 to 4 orders of magnitude lower than those of GCs.

In this study, we perform GC orbital integrations for our sample of 198 MW-like galaxies across cosmic time, within time-evolving FDM MW potentials, for four different FDM particle masses: $m_{22} = 0.3$, 1, 7, and 30, using the publicly available code \texttt{galpy}\footnote{Available at \url{https://github.com/jobovy/galpy}} \citep{Bovy15}. We also account for mass loss of our 7709 GCs for each DM model. At each MW potential update, GC mass loss is computed using the model of \cite{Kruijssen11}, which includes contributions from stellar evolution, two-body relaxation, and tidal shocks. Our orbital integration time resolution is set to 2 Myr (500 steps per Gyr), using the fast C integrator \texttt{dop853\_c} implemented in \texttt{galpy}. This is an explicit Runge–Kutta method of order 8(5,3), which offers high accuracy and efficiency for evolving complex dynamical systems. Regarding computational performance, orbital integrations from $z = 2$ to 0 for $m_{22} = 1$ take approximately 6 CPU hours.

\section{FDM confined or expanded-orbit mechanism}

Figure~\ref{fig05} illustrates how the orbits of in-situ GCs evolve over time depending on the DM model, based on two example MWs. The initial distribution is shown in the normalized energy–angular momentum space at $z=2$ in the left panels. We highlight that the orbital decay of the clusters, in the absence of dynamical friction, is caused by the time evolution of the MW potential.

Compared to the orbits in CDM, those in FDM with $m_{22} = 0.3$ are significantly contracted from the very first Gyr. A persistent trapping below $r < 2$--$3$ kpc is observed after only a few Gyr, due to the steep slope in the $E(L_z)$ curve for the FDM $m_{22} = 0.3$ model (see left panels of Figure~\ref{fig05}), which may seem counter-intuitive. Indeed, FDM with $m_{22} = 0.3$ exhibits a large, low-density DM core, in contrast to the cuspy NFW profile (see Figure~\ref{fig01}). However, the steepness of the $E(L_z)$ relation is explained by the dominant baryonic contribution in the central regions of the galaxy. The stellar profile, modeled with a highly concentrated Hernquist potential, generates a deep central potential well that strongly affects low-radius orbits, inducing a rapid variation of energy with angular momentum. This steep slope means that a GC must gain significantly more energy to increase its angular momentum and move to a wider orbit. In other words, the energy barrier to escape the central region is harder to overcome in FDM ($m_{22} = 0.3$): GCs are more “trapped” than in CDM. We refer to this phenomenon as orbital confinement throughout the paper. At larger radii, the baryonic density drops rapidly, allowing DM to dominate. In the case of FDM with $m_{22} = 0.3$, the extended and diffuse core results in a gradual flattening of the $E(L_z)$ curve (see Figure~\ref{fig05}). We also note that the orbital space in the $E$-$L_{z}$ diagram (region of bound and stable orbits) is relatively restricted, despite the presence of a deep baryonic potential. This is further reinforced by the fact that the total stellar mass is much smaller than the DM mass. Although baryons dominate locally, they cannot, on their own, sustain a deep and extended gravitational potential.

In contrast, in FDM with $m_{22} = 7.0$, GCs can reach galactocentric distances that are even larger than in CDM, as shown in right panels of Figure~\ref{fig05}. As $m_{22}$ increases, the DM core becomes more compact and denser, altering the gravitational balance (see Figure~\ref{fig01}). As a result, the relative contribution of baryons to the total potential decreases, and DM takes over a larger fraction of the galaxy. The global potential thus becomes deeper and more extended, allowing for a wider range of accessible orbital energies. This enables the existence of stable orbits over a broader range of angular momenta, even far from the baryonic center. Consequently, the accessible area in the $(E, L_z)$ plane expands. In FDM with $m_{22} = 7.0$, GC orbits are more extended than in CDM. We refer to this as orbital expansion.

In summary, at low $m_{22}$, baryons dominate a short and steep central potential, which confines the orbital space to a narrow region. At higher $m_{22}$, DM becomes both compact and globally dominant, supporting a deeper and more extended gravitational potential that allows for a wider variety of stable orbits. A natural consequence of these orbital behaviors is that GCs confined near the center are much more susceptible to tidal effects and, ultimately, to disruption unlike those that can migrate to larger galactocentric distances and remain protected from strong tidal forces.

\begin{figure}
    \centering
    \includegraphics[width=\linewidth]{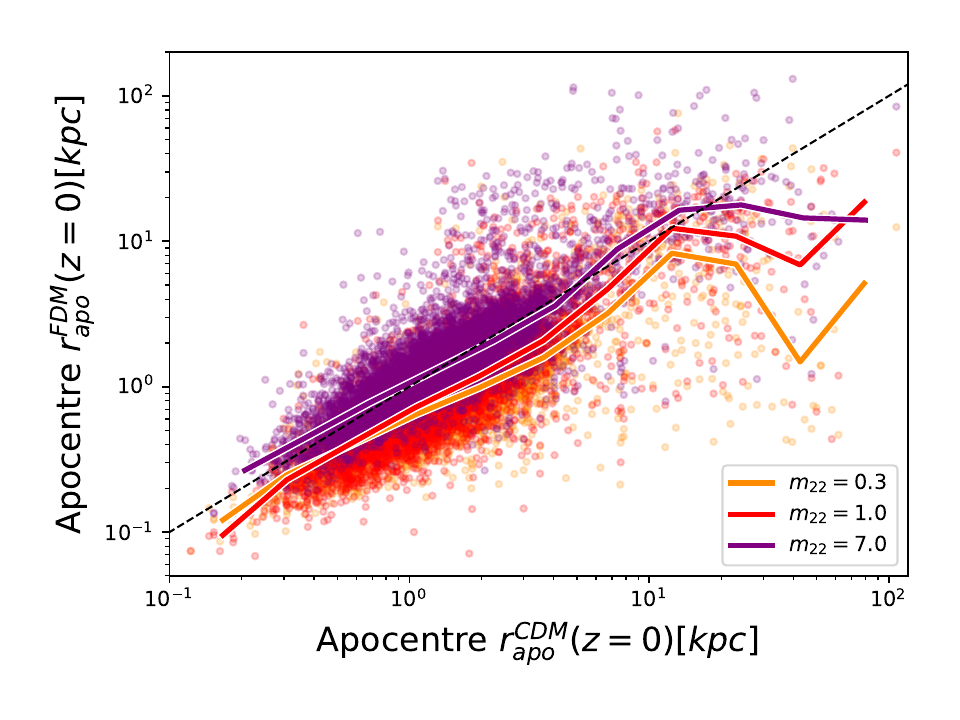}
    \caption{Last apocenter in FDM as a function of the last apocenter in CDM for in-situ GCs across our entire MW sample for different DM models. The three colors represent different FDM models, with $m_{22} = 0.3$ (orange), $1.0$ (red), and $7.0$ (violet). The lines indicate the median values.}
    \label{fig06}
\end{figure}

To verify these two regimes of orbital behavior across our entire MW sample, Figure~\ref{fig06} compares, for the 7709 simulated in-situ GCs, the last orbital apocenter reached in an FDM MW to that reached in CDM. The apocentre directly captures the effects of orbital contraction and expansion. The downward deviation of the curves for $m_{22}=0.3$ and 1 from the diagonal shows that, for an equivalent initial orbit in the $E$-$L_{z}$ space, a GC evolving in an FDM MW remains more spatially confined than in CDM. This effect is more pronounced at lower $m_{22}$ values. For $m_{22} = 0.3$, apocenters in FDM are up to a factor of 3–5 smaller than in CDM over a wide range of orbits. Conversely, for $m_{22} = 7.0$, the behavior is similar to CDM, with deviations at large distances where the apocenter first increases before decreasing (see Figure~\ref{fig06}). We also verified that for FDM with $m_{22} = 30$, there is a pronounced increase in apocenter in the central region that diminishes at larger distances. The reduction or increase of the apocenter in FDM as a function of $m_{22}$ confirms that clusters enter two distinct regimes. Either they fail to explore the outer regions of MW-like galaxies, which limits their orbital dispersion and increases their dynamical vulnerability, or they can reach more distant regions, thereby enhancing their chances of survival within the galaxy.

\section{FDM impact on the in-situ GC population at $z=0$}

Our first investigations have shown that FDM not only alters halo structure through core formation, but also profoundly modifies the orbital dynamics of clusters by either restricting or enhancing the radial extent of their orbits. This confinement for $m_{22} \leq 1$ or expansion for $m_{22} > 1$ appears as a clear dynamical signature of the FDM effect. This motivates us to examine whether these orbital behaviors have significant consequences on the GC distribution at $z = 0$, which could not only distinguish FDM from CDM, but also constrain the free parameter $m_{22}$ through observations. We anticipate that orbital confinement or expansion will directly affect both the survivability and the spatial distribution of the in-situ GC population. For this reason, we have selected specific metrics to quantify these effects as a function of the stellar mass of the MWs in our sample.

\subsection{GC mass distribution}

\begin{figure}
    \centering
    \includegraphics[width=\linewidth]{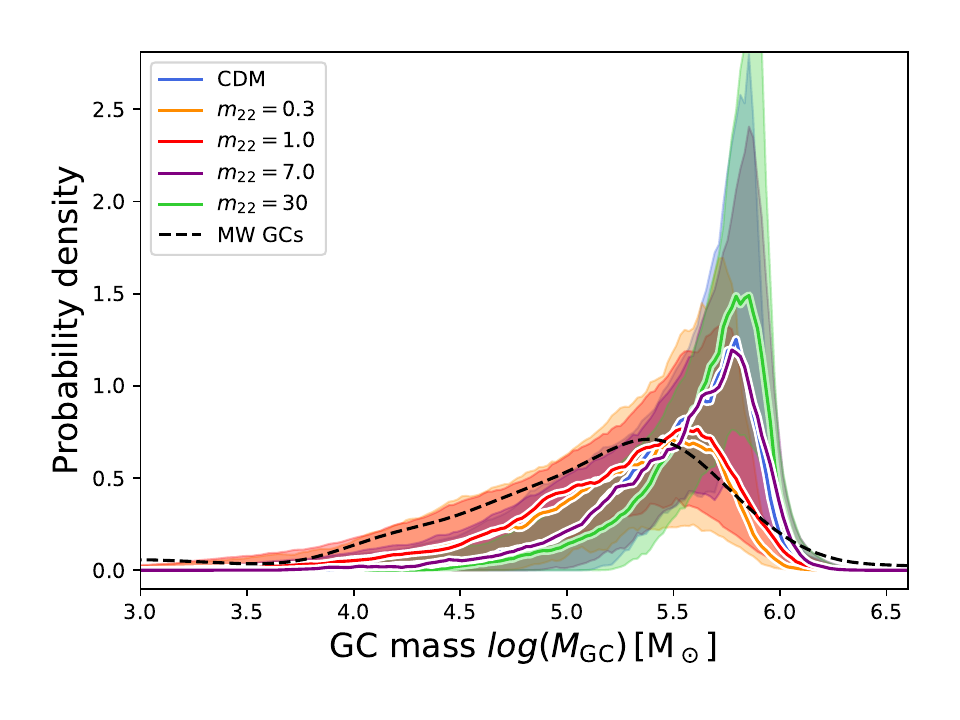}
    \caption{Probability density of GC mass at $z=0$ for different DM models. Solid lines show the median trends, with shaded regions indicating the 25th and 75th percentiles. The black dashed line corresponds to the observed GC population in the MW, although it includes both in-situ and ex-situ clusters.}
    \label{fig07}
\end{figure}

Figure~\ref{fig07} shows the probability density distribution of in-situ GC masses at $z = 0$ for different DM models.  For all FDM models with $m_{22} \leq 1$, the distribution is tightly clustered around $\log(M_{\mathrm{GC}}/M_\odot) \approx 5.6$. In contrast, CDM and FDM models with $m_{22} > 1$ exhibit narrower and more peaked distributions centered around $\log(M_{\mathrm{GC}}/M_\odot) \approx 5.8$. Moreover, the models differ significantly in the low-mass end of the distribution. CDM and FDM models with $m_{22} \geq 7$ show a clear deficit of low-mass clusters ($\log(M_{\mathrm{GC}}/M_\odot) < 5$). On the other hand, FDM models with lower $m_{22}$ values (especially $m_{22} = 1.0$ and $0.3$) produce broader distributions, with a significant number of low-mass clusters. This behavior is explained by the fact that in these models, GCs are more vulnerable to tidal effects, particularly when trapped in the central regions of the galactic potential, which is a typical outcome for FDM with low $m_{22}$. In contrast, the $m_{22} = 30$ model shows the highest peak in the distribution, as clusters on more extended orbits can avoid the strong tidal field of the central regions. 

Thus, from an observational perspective, a narrow mass distribution dominated by massive clusters would be difficult to reconcile with low-$m_{22}$ FDM models, while an excess of surviving low-mass clusters would be a strong indicator in favor of fuzzy DM with low $m_{22}$. Comparing these predictions to the observed mass distribution in the MW could therefore offer a direct test of the nature of DM once the in-situ and ex-situ GC populations are properly separated. This is challenging, however, because the observed population is a mixture of both in-situ and ex-situ GCs \citep{Pagnini23,Boldrini25}.

Furthermore, our current results suggest that reproducing the observed GC mass distribution in the MW requires distinct properties for the ex-situ population depending on the value of $m_{22}$. For low $m_{22}$, ex-situ clusters must be predominantly low-mass, with $\log(M_{\mathrm{GC}}/M_\odot) \lesssim 5.4$, in order to compensate for the lack of light in-situ GCs. In contrast, for high $m_{22}$, the ex-situ population must fill the deficit across the entire mass range, from $\log(M_{\mathrm{GC}}/M_\odot) \approx 4.5$ to $6.2$, and therefore must dominate the total GC population.

\begin{figure*}
    \centering
    \includegraphics[width=\linewidth]{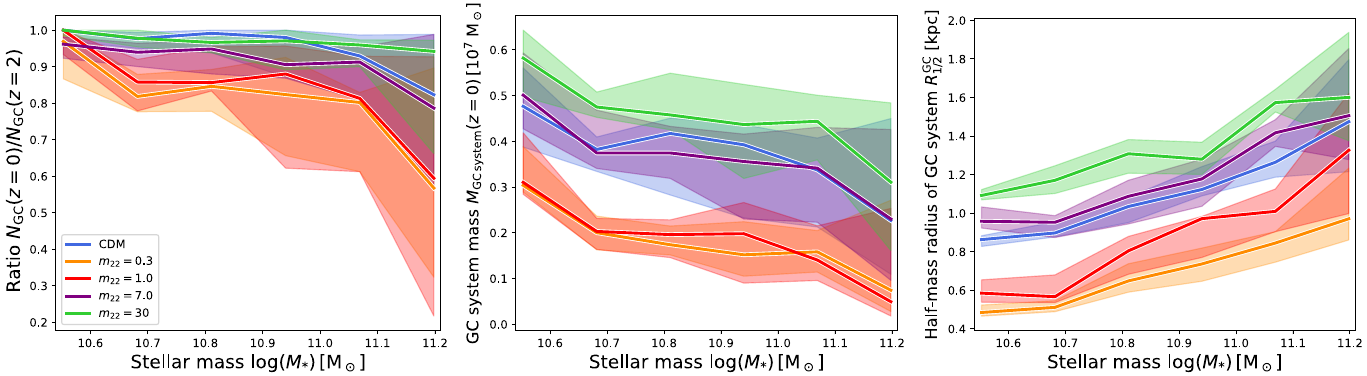}
    \caption{FDM impacts at $z=0$: Left panel: Ratio of the number of in-situ GCs between $z = 2$ and $z = 0$ as a function of MW stellar mass at $z=0$. Middle panel: Total mass of the in-situ GC system at $z=0$ as a function of MW stellar mass. Right panel: Half-mass radius of the in-situ GC system at $z=0$ as a function of MW stellar mass. Each panel compares our different FDM models ($m_{22} = 0.3$, $1.0$, $7.0$, and $30$) to CDM for the 198 MW-like galaxies. Solid lines and their associated shaded regions represent the median and 68$\%$ confidence intervals within a moving bin along stellar mass. The confidence intervals are estimated from 1000 bootstrap realizations, each using 95$\%$ of the original sample size.} 
    \label{fig08}
\end{figure*}

\subsection{In-situ GC survival, mass, and extent in FDM}

We analyze the impact of FDM by measuring the number of surviving clusters, the total mass, and the half-mass radius of the in-situ GC system at $z=0$, relative to CDM. The left panel of Figure~\ref{fig08} shows the fraction of surviving in-situ GCs, i.e. those not destroyed by tidal effects between $z = 2$ and $z = 0$ in various MW potentials from CDM and FDM models with $m_{22} = 0.3$, $1.0$, $7.0$, and $30$. In the CDM model, the survival fraction of in-situ GCs remains relatively high, exceeding 95$\%$ for galaxies with stellar masses below $\log(M_*/M_\odot) \approx 11$, and dropping below 90$\%$ for more massive systems. FDM models with high $m_{22}$ values (e.g., $m_{22} = 7$ or $30$) display similar behavior to CDM. In contrast, for FDM with $m_{22} = 1$, the survival fraction drops below 90$\%$ already at $\log(M_*/M_\odot) = 10.7$, and falls below 80$\%$ for $\log(M_*/M_\odot) > 11.0$. The most extreme case occurs for $m_{22} = 0.3$, where the fraction is already below 85$\%$ in low-mass galaxies and drops below 70$\%$ in the most massive ones (see left panel of Figure~\ref{fig08}). This result indicates that more massive galaxies destroy in-situ GCs more efficiently. Indeed, the stronger tidal forces in the central regions of more massive systems increase the likelihood of disruption. In addition, FDM models with low $m_{22}$ amplify this effect by confining clusters more tightly to the central regions, where tidal forces are most destructive.

The middle panel of Figure~\ref{fig08} depicts the total mass of the in-situ GC system at $z = 0$ as a function of the MW stellar mass. Overall, in all models, the total GC mass tends to decrease with increasing stellar mass. In the CDM model, low-mass galaxies with $\log(M_*/M_\odot) < 10.7$ typically host in-situ GC systems more massive than $4 \times 10^6\ M_\odot$, whereas for $\log(M_*/M_\odot) > 11.0$, the GC mass drops below $4 \times 10^6\ M_\odot$. The FDM model with $m_{22} = 30$ exhibits systematically higher GC masses than CDM, reaching $\sim 6 \times 10^6\ M_\odot$ in low-mass galaxies and remaining above $4 \times 10^6\ M_\odot$ even in the most massive systems. Conversely, FDM models with $m_{22} = 0.3$ or $1$ show much lower in-situ GC masses: below $3 \times 10^6\ M_\odot$ in small galaxies, and falling under $2 \times 10^6\ M_\odot$ for the most massive ones (see Figure~\ref{fig08}). This confirms that the destruction of in-situ GCs is more severe in more massive halos, and that low-$m_{22}$ FDM models further enhance this destruction, likely due to the prolonged confinement of clusters in the central regions of the potential.

The right panel of Figure~\ref{fig08} presents the evolution of the half-mass radius of the in-situ GC system at $z = 0$, as a function of the stellar mass. A general increasing trend is observed across all models. More massive galaxies host spatially more extended GC systems. However, the different DM models predict systematically different sizes. FDM models with high boson mass ($m_{22} = 7$ or $30$) produce more extended systems than CDM, reaching approximately $1.2$ to $1.5$ kpc for $\log(M_*/M_\odot) > 11.0$, as shown in Figure~\ref{fig08}. The CDM model shows intermediate sizes, with half-mass radii ranging from about $0.9$ kpc for $\log(M_*/M_\odot) < 10.7$ to $\sim 1.5$ kpc in the most massive galaxies. In contrast, low-$m_{22}$ FDM models ($m_{22} = 0.3$ or $1.0$) produce significantly more compact systems. For instance, in the $m_{22} = 0.3$ model, the half-mass radius never exceeds $1$ kpc across the entire stellar mass range (see Figure~\ref{fig08}). This behavior arises because low-$m_{22}$ FDM MWs tend to confine GCs more tightly within the central regions of the galaxy, reducing the spatial extent of the GC system at $z = 0$. Conversely, for high $m_{22}$ values, the opposite effect is observed. The orbital space accessible in the $E$-$L_{z}$ plane widens, allowing GCs to reach more distant regions than in CDM. This implies that, observationally, a compact in-situ GC system could be a signature of an FDM model with low $m_{22}$, while more extended systems are consistent with high-$m_{22}$ FDM models.

In a universe dominated by FDM, we identify three regimes for the in-situ GC population, depending on the value of $m_{22}$:
\begin{itemize}
    \item For $m_{22} < 7$: the GC population consists of fewer and individually less massive clusters. The resulting system is therefore less massive and more spatially concentrated than in the CDM scenario.
    \item For $m_{22} \sim 7$: the properties of the GC systems are similar to those obtained in the CDM model, both in terms of total mass and spatial distribution.
    \item For $m_{22} > 7$: the population includes a larger number of surviving clusters, generally more massive. The resulting system is thus more massive and more extended than in the CDM scenario.
\end{itemize}

\section{Extension to self-interacting and warm dark matter theories}

Over the past decades, a wide range of alternative DM models have been proposed. Among them, three major classes have primarily been explored through simulations: warm DM, self-interacting DM and FDM. Our approach focuses on the impact of the DM distribution modified by FDM on the dynamics of in-situ GCs. Our study shows that deviations from the CDM scenario, combined with baryonic effects, significantly modify the orbital phase space accessible to GCs, resulting in clear signatures in the energy–angular momentum $E$-$L_{z}$ diagrams. In principle, a similar analysis could be carried out for SIDM and WDM models, since these frameworks also predict the formation of DM cores whose properties depend on a free DM parameter. However, unlike FDM, where redshift-dependent evolution equations for halo structure exist (see Equations~\eqref{rhoFDM} and~\eqref{rcFDM}), no such time-evolving description is currently available for SIDM and WDM. As a result, it is presently difficult to construct cosmologically consistent models in these scenarios. We therefore propose to extrapolate the results obtained in the FDM case to SIDM and WDM, based on a comparison of the $E$-$L_{z}$ diagrams at $z=0$, where the dynamical properties of these models are reasonably well understood. The goal is to motivate future dedicated simulations in these alternative DM theories, despite their high computational cost, by first reducing the dimensionality of the relevant parameter space (free theory parameter, galaxy mass range). Indeed, the main limitation of these alternative models lies in the introduction of an additional free parameter such as particle mass (FDM, WDM) or interaction cross-section (SIDM) compared to CDM, which must be explored systematically.

\subsection{Predictions for FDM, SIDM and WDM}

To enable a more direct comparison between alternative DM theories and to predict their impact on the dynamics of in-situ GCs, we introduce the following metric: the relative orbital index, $I_{\rm orb}$. It corresponds to the normalized area of the bound region in the $E$–$L_z$ space, relative to the CDM case. This area represents the region of allowed orbits below a fixed energy threshold $E_{\rm th}$ in the $E(L_z)$ diagram. The threshold energy is defined as the energy of a circular orbit in the CDM model at the scale radius $r_s$. The relative orbital index is then defined as:
\begin{equation}
    I_{\rm orb} = A(p_{\rm DM}) / A_{\rm CDM},
\label{iorb}
\end{equation}
where $A(p_{\rm DM})$ is the orbital area for the alternative DM model characterized by the parameter $p_{\rm DM}$, and $A_{\rm CDM}$ is the corresponding area in CDM. This calculation is performed for two representative halo masses: $10^9\ M_\odot$ (dwarf galaxies, dotted lines) and $10^{12}\ M_\odot$ (MW galaxies, solid lines), across the WDM, SIDM, and FDM theories.

\begin{figure}
    \centering
    \includegraphics[width=\linewidth]{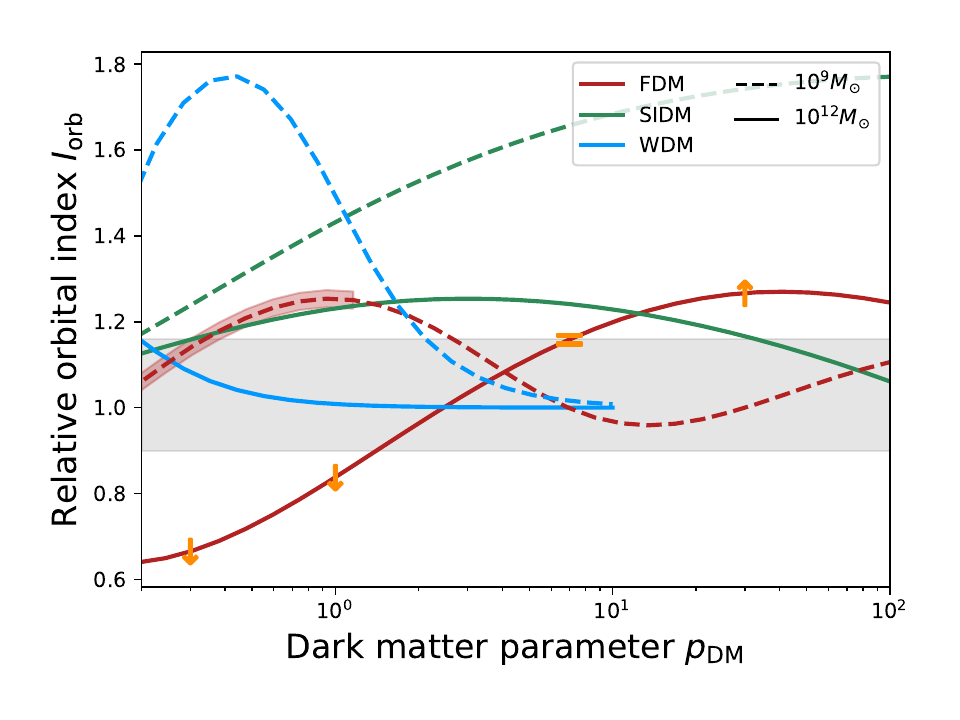}
    \caption{Relative orbital index as a function of the DM parameter $p_{\mathrm{DM}}$ for FDM, WDM, and SIDM, shown for a $10^{9}\ M_\odot$ dwarf-like galaxy (dotted lines) and a $10^{12}\ M_\odot$ MW-like galaxy (solid lines). The relative orbital index is defined in Equation~\eqref{iorb}. The parameter $p_{\mathrm{DM}}$, characterizing the strength of the deviations from CDM, corresponds to the FDM particle mass $m_{22}$, the WDM particle mass $m_{\nu}$ (in keV), and the SIDM cross-section $\sigma_m$ (in cm$^2$/g). Simulation results for FDM with $m_{22} = 0.3$, $1$, $7$, and $30$ are shown as orange markers. Downward (upward) arrows indicate orbital confinement (expansion) of in-situ GCs relative to CDM. An equal sign indicates that we recover a CDM-like orbital behavior at $m_{22} = 7$. The red shaded region for the FDM dwarf-like galaxy marks the regime where FDM dynamical friction significantly deviates from that in CDM.}
    \label{fig010}
\end{figure}

Figure~\ref{fig010} shows how the relative orbital index varies as a function of the DM parameter $p_{\rm DM}$, which captures the strength of the deviation from the CDM case. According to our results, values of $I_{\rm orb} > 1$ correspond to an enlarged phase space, allowing a greater number of orbital configurations and favoring orbital expansion. Conversely, $I_{\rm orb} < 1$ indicates a reduced phase space, limiting the available orbital configurations and leading to orbital confinement. In the case of FDM (red curves), the DM parameter is the boson mass $m_{22}$. Moving leftward on the horizontal axis thus corresponds to lighter bosons, which leads to an increase in the DM core size, as described by Equation~\eqref{rcFDM}. For SIDM (green curves), the relevant parameter is the self-interaction cross-section per unit mass $\sigma/m$. Moving rightward corresponds to a higher interaction rate, which results in a larger DM core (see Figure~\ref{fig09}). In the case of WDM (blue curves), the DM parameter is proportional to the WDM particle mass $m_\nu$. Rightward shifts correspond to lighter particles, which likewise increase the core size (see Figure~\ref{fig09}).

\subsection{MW regime}

For a MW-like halo, the FDM, WDM, and SIDM curves display clearly distinct behaviors. The orbital index in Figure~\ref{fig010} highlights the three regimes identified previously for FDM. We overplot the results of our FDM simulations for $m_{22} = 0.3$, $1$, and $7$ using orange markers. Downward (upward) arrows indicate orbital confinement (expansion), while the equal sign denotes behavior similar to CDM. Based on our simulation results, we define a gray band representing the parameter space where orbital behavior is indistinguishable from CDM according to our specific metrics shown in Figure~\ref{fig08}. Using this calibration, we predict the orbital behavior of in-situ GCs in WDM and SIDM in MW-like galaxies. For WDM, the relative orbital index remains within the gray band across the full range of $p_{\rm DM}$, implying that the GC orbital dynamics are essentially indistinguishable from CDM. By contrast, in SIDM with $\sigma_m = 1$--$10$ cm$^2$/g, the orbital index rises above the gray band with an amplitude comparable to that of the FDM simulation with $m_{22} = 30$ (see Figure~\ref{fig010}). This suggests that GCs in SIDM may reach more extended orbits, with consequences for the in-situ GC population at $z=0$. In summary, only FDM and SIDM are expected to leave a potential dynamical imprint on the in-situ GC population in MW-like galaxies. Unlike SIDM, which only induces orbital expansion relative to CDM, FDM can produce either orbital confinement or expansion depending on the value of its free parameter, $m_{22}$.

\subsection{Dwarf galaxy regime}

Considering the dwarf galaxy regime also allows us to evaluate the impact of halo mass in each alternative DM theory. For FDM and WDM models, the size of the DM core increases as the halo mass decreases, while the opposite trend is observed in the SIDM scenario, where the core size is reduced in low-mass halos. In this low-mass regime, dynamical friction becomes effective again, in contrast to what is observed in MW-like galaxies. Indeed, the characteristic timescale for significantly altering a GC's apocenter through dynamical friction is inversely proportional to the mass ratio between the GC and its host galaxy, which is smaller in dwarf galaxies. Compared to CDM, dynamical friction is expected to be weaker due to the presence of a DM core, which lowers the central density. In addition, theory-specific effects may further reduce friction: in FDM, intrinsic density fluctuations can suppress dynamical friction, while in SIDM, self-interactions have a similar effect. These effects can be quantified by the quantum Mach number (FDM) and the Knudsen number (SIDM), which characterize regimes where dynamical friction is significantly reduced.  
In contrast, aside from the presence of a core, dynamical friction in WDM remains broadly similar to CDM. Finally, it is worth noting that in dwarf galaxies, heating of GCs due to FDM quantum fluctuations becomes even less efficient than in MW-like systems, owing to the $t_{\mathrm{dis}} \sim \rho^{-2}$ dependence, which leads to much longer disruption times in low-density environments.

To model a dwarf galaxy, we considered a $10^{9}\ M_\odot$ DM halo and a stellar distribution described by a Hernquist profile. The stellar mass and half-mass radius were estimated by fitting the stellar masses and sizes of satellite galaxies of MW at $z = 0$ in TNG50, as identified by \cite{2021MNRAS.507.4211E}. According to Figure~\ref{fig010}, GCs in FDM dwarf galaxies behave similarly to those in CDM, except in the range $m_{22} = 0.4$–$2$. In this boson mass interval, the orbital behavior is expected to mimic the expansion observed in our FDM simulation with $m_{22} = 30$. Additionally, we highlight that FDM dynamical friction becomes non-negligible in this range (see the red shaded region in Figure~\ref{fig010}). To quantify this effect, we computed the quantum Mach number (see Equation~\eqref{MQN}) for a circular orbit at the scale radius. A threshold of $\mathcal{M}_Q = 100$ was adopted to identify the $\mathcal{M}_Q \gg 1$ regime, where classical dynamical friction is no longer valid. FDM progressively suppresses dynamical friction as $m_{22}$ decreases. Overall, for $m_{22} = 0.4$–$1$, we predict a modified orbital behavior compared to CDM, due to the combined effects of orbital expansion and reduced dynamical friction. These mechanisms are expected to enhance the survival of GCs in dwarf galaxies relative to the CDM case.

For WDM and SIDM, the orbital behavior changes significantly compared to the MW regime (see dashed lines in Figure~\ref{fig010}). For SIDM, the orbital index increases with $\sigma_{m}$, reaching values up to $\sim$1.8. This indicates that as the DM core grows in size, the dwarf galaxy becomes increasingly favorable to a wider range of orbits. As a result, SIDM dwarf galaxies are expected to host GC systems that are spatially more extended and retain masses close to their initial values, since tidal forces are weaker at large distances. Regarding dynamical friction, the Knudsen number $K_{n}$ \citep{2011MNRAS.415.1125K}, which ranges from 19 to 311 for $\sigma_{m} = 100$–$0.2\ \mathrm{cm^2/g}$, suggests that the system remains in the classical regime, where gravity dominates over scattering if $K_{n} \gg 1$ \citep{2024A&A...690A.299F}. Despite a dynamical friction comparable to CDM, our analysis predicts that the amplitude of SIDM's impact on GCs is generally stronger than in the FDM scenario across nearly the entire parameter space $p_{\mathrm{DM}}$. In the case of WDM, the orbital index in dwarf galaxies exceeds 1.5 for $m_{\nu} < 1$ keV and rapidly decreases toward the grey band for higher particle masses (see Figure~\ref{fig010}). This means that for large DM cores in dwarfs, WDM can induce significant orbital expansion relative to CDM when $m_{\nu} < 1$ keV — with an amplitude even greater than in our FDM simulation with $m_{22} = 30$.

In summary, except within a very narrow range of $m_{22}$ values, no significant deviation is expected between FDM and CDM in dwarf galaxies. In contrast, we anticipate substantial differences in the GC population of dwarfs for WDM when $m_{\nu} < 1$ keV and for SIDM across nearly the entire range of $\sigma_{m}$. Based on these predictions, we expect the ex-situ GC population — that is, clusters accreted during satellite mergers — to be affected differently in the WDM and SIDM models compared to CDM. Outside of regimes where the effects are negligible and orbital distributions remain close to CDM expectations, satellite accretion onto MW-like galaxies should result in an in-situ/ex-situ mixture whose dynamical properties depend strongly on the underlying DM model. This leads to three distinct scenarios:
\begin{itemize}
    \item FDM: The GC population of a MW-like galaxy is composed of a perturbed in-situ component, whose orbits may be either confined or extended depending on the value of $m_{22}$, and an ex-situ component whose dynamics remains globally similar to the CDM case.
    \item WDM: The population is dominated by an ex-situ component strongly affected by a more pronounced orbital expansion mechanism than in the FDM case, while the in-situ component remains very close to the CDM prediction.
    \item SIDM: Both the in-situ and ex-situ components undergo orbital expansion compared to CDM, driven by the self-interactions of DM.
\end{itemize}

\subsection{Constraints on the DM free parameter}

Here, we discuss the implications of our results in light of the existing constraints on the free DM parameters of the three theories. According to Figure~\ref{fig010}, galactic-scale constraints favor the confinement of GCs, whereas large-scale constraints favor orbital expansion of clusters in MW-like galaxies. For dwarf galaxies, only galactic constraints are consistent with a modification relative to CDM in the dynamics of GCs.

In order to alleviate the cusp-core problem while also satisfying observational constraints across different mass scales, SIDM models must feature a velocity-dependent cross-section that decreases with halo mass \citep{2016PhRvL.116d1302K}. A compilation of such constraints is presented in \cite{2024MNRAS.529.2327F}. \cite{2021MNRAS.503..920C} established that dwarf galaxies ($10^9\,M_\odot$) allow for SIDM cross-sections in the range $\sigma/m \sim 30$--$100\,{\rm cm^2/g}$. In contrast, for MW-like galaxies ($10^{12}\,M_\odot$), the allowed range is more constrained, typically $\sigma/m \sim 1$--$10\,{\rm cm^2/g}$ \citep{2025MNRAS.536.3338C}. These results confirm the necessity of adopting a velocity-dependent self-interaction cross-section in SIDM models to remain consistent with observations across different mass scales. According to Figure~\ref{fig010}, these constraints define the specific regimes where the orbital dynamics of GCs are expected to deviate most significantly from CDM predictions, both in dwarfs and MW-like galaxies. Therefore, GCs remain a viable probe of the nature of DM.

For WDM, an upper limit of 2 keV ($2\sigma$) has been imposed by Lyman-$\alpha$ forest observations \citep{2005PhRvD..71f3534V,2013MNRAS.432.3218D,2014PASA...31....6M}. According to Figure~\ref{fig010}, this constraint corresponds to a regime in which the orbital dynamics of GCs in WDM are expected to be very similar to those predicted by the CDM model. However, Lyman-$\alpha$ forest constraints rely on sensitive assumptions about the intergalactic medium as well as the resolution of the simulations used \citep{2022Univ....8...76P}. A more complete modelling of the physical properties of WDM, particularly the role of quantum pressure, could allow for a more nuanced and accurate interpretation of these results \citep{2022Univ....8...76P}.

\section{Conclusion}

In this paper, we presented the first systematic study of the dynamics of in-situ GCs in MW-like galaxies embedded in FDM halos. Our approach combined cosmological mass accretion histories from the TNG50 simulation with time-dependent FDM potentials and dedicated orbital integrations, allowing us to follow the long-term evolution of cluster populations from $z=2$ to $z=0$. This strategy provided a computationally efficient yet physically motivated framework to explore the dynamical consequences of alternative DM models on GC systems. Our analysis reveals that the orbital structure and survivability of in-situ GCs depend sensitively on the FDM boson mass $m_{22}$. We identify three distinct regimes: Low-mass regime ($m_{22} < 7$): baryons dominate the inner potential, confining clusters into compact and tidally fragile configurations. The resulting systems are less massive and spatially concentrated compared to CDM, with a broader GC mass distribution extending to lower masses; Intermediate regime ($m_{22} \sim 7$): the orbital space, mass distribution, and spatial extent of in-situ clusters closely match those found in CDM, producing nearly indistinguishable systems at $z=0$.; High-mass regime ($m_{22} > 7$): the DM halo becomes more compact and globally dominant, supporting a deeper and more extended potential that allows clusters to explore wider orbits. Surviving systems are more massive, more extended, and depleted in low-mass clusters relative to CDM. These dynamical regimes reflect a fundamental competition between baryonic and DM components, which directly imprints on observable quantities such as the survival fraction, total mass, and half-mass radius of in-situ GC populations. The predicted differences are of sufficient amplitude to offer a new avenue to constrain the nature of DM, provided that in-situ and ex-situ clusters can be observationally disentangled. Future work should focus on incorporating the ex-situ population, accreted during satellite mergers, by explicitly accounting for hierarchical assembly processes and adopting more realistic initial conditions for GCs, as proposed in \cite{Boldrini25}. This comprehensive model could then be constrained using the present-day mass function of MW GCs provided by \cite{Baum18}. Such a calibration would yield crucial implications for the demographics of the high-redshift cluster population in both CDM and FDM scenarios, ultimately paving the way for stringent tests against forthcoming JWST observations \citep{2025MNRAS.537.2535C}.

Extending our framework to SIDM and WDM models shows that these scenarios can also alter orbital phase space, but with distinct signatures. In particular, SIDM halos tend to induce orbital expansion without confinement, while WDM halos leave GC dynamics largely unchanged relative to CDM at MW scales. This highlights the confinement orbital mechanism as a unique dynamical signature of FDM. Moreover, our generalized orbital index provides a compact metric to compare alternative DM scenarios across different mass scales. Unlike FDM, where redshift-dependent evolution equations for halo structure exist, no such time-evolving description is currently available for SIDM and WDM. However, these models are numerically less demanding to simulate than FDM, and large-volume cosmological runs with box sizes comparable to TNG50 already exist (e.g. Darkium simulation for SIDM \citep{Darkium}). In addition, the treatment of dynamical friction in SIDM and WDM does not require a specific formulation, as it closely resembles that in CDM.

These results underscore the potential of GC systems as dynamical probes of
DM physics. The ongoing Euclid mission will be able to provide the half-mass radius of GC population across its unprecedented coverage of galaxies, spanning stellar masses from $10^{9}$ M$_\odot$ (dwarfs) to $10^{12}$ M$_\odot$ (MW-like systems) in the nearby Universe ($< 100$~Mpc), by detecting half a million extragalactic GCs \citep{2021sf2a.conf..447L,Voggel25,Sai2025V1}. This wealth of data will enable statistical tests of the regimes identified here, in particular the predicted compactness of in-situ systems at low $m_{22}$ and their expansion at high $m_{22}$. A finer metric to normalize the GC distribution could be the half-mass radius of field stars in MW-like galaxies detected by Euclid. Indeed, the dynamical heating of stars, induced by exotic DM properties such as self-interactions (SIDM) or density fluctuations (FDM), affects the surrounding gas and stars, promoting the formation of more extended galaxies \citep{2025MNRAS.536.3338C}. Unlike GCs, the FDM heating timescale is shorter for field stars. The half-mass radius ratio between GCs and stars could therefore be crucial to constrain FDM, given the in-situ stellar populations identified in MW-like galaxies. In the case of SIDM, this behavior arises where baryons dominate the central gravitational potential \citep{2025MNRAS.536.3338C}, which occurs only for low $m_{22}$ in FDM. If similar heating beahavior occurs in FDM, we expect the GC-to-stellar half-mass radius ratio to be generally lower for low $m_{22}$ compared to CDM. Conversely, for high $m_{22}$ where DM dominates the central region, the opposite behavior is expected: the ratio should increase relative to CDM, as the GC distribution is more extended while the stellar half-mass radius remains similar or larger than in CDM.

\section{Data Availability}

The data underlying this article is available through reasonable request to the author. The code and GC data will be available at the following URL: \href{https://github.com/Blackholan}{https://github.com/Blackholan.}

\begin{acknowledgements} 

PB acknowledges funding from the CNES post-doctoral fellowship program. This
work was also supported by CNES, focused on the Gaia mission. PB and PDM are grateful to the "Action Thématique de Cosmologie et Galaxies (ATCG), Programme National ASTRO of the INSU (Institut National des Sciences de l'Univers) for supporting this research, in the framework of the project "Coevolution of globular clusters and dwarf galaxies, in the context of hierarchical galaxy formation: from the Milky Way to the nearby Universe" (PI: A. Lançon).

\end{acknowledgements}

\bibliography{src}

\appendix

\section{FDM heating disruption time}

\begin{figure}
    \centering
    \includegraphics[width=\linewidth]{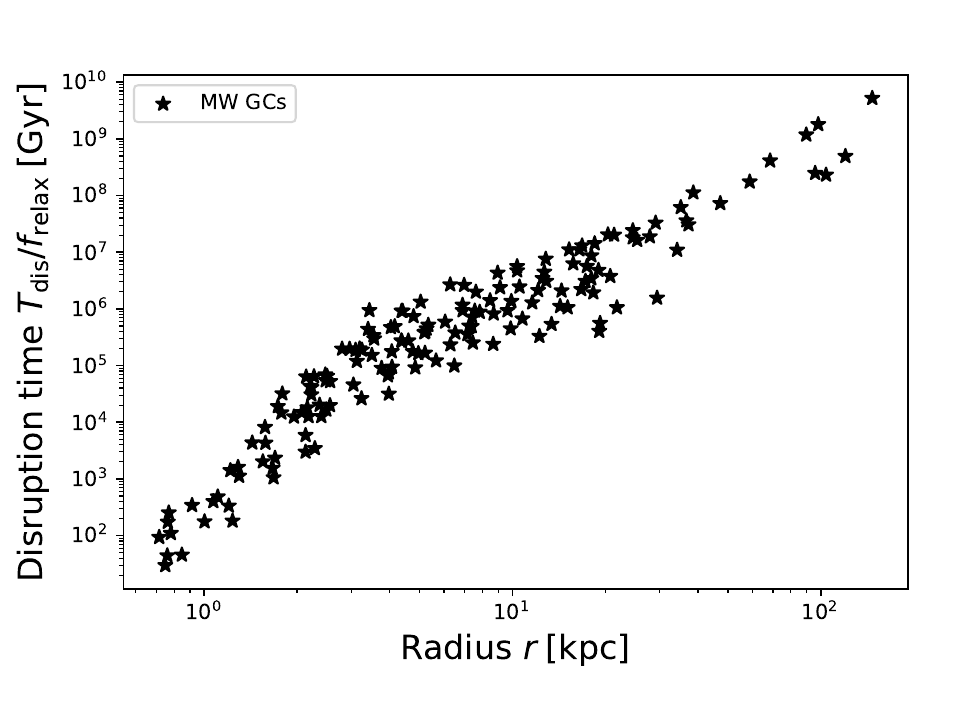}
    \caption{Disruption time due to FDM fluctuation-induced heating as a function of orbital distance for the 171 GCs in the MW, as provided by Gaia \citep{Vasiliev21}. The disruption time was computed using Equation~\eqref{tdis}, assuming $m_{22} = 0.3$ for an FDM halo with a total mass of $10^{12}~M_{\odot}$. The disruption process is exceedingly slow for MW GCs, even in the central region, with timescales ranging from 30 to 400 Gyr within the inner 1 kpc. In addition, the timescale of the process increases with the boson mass $m_{22}$}
    \label{fig04}
\end{figure}

\section{Structure of DM halos and on the orbital phase space of FDM, SIDM, WDM models}

All plots in Figure~\ref{fig09} correspond to an example MW with $M_{\mathrm{halo}} = 10^{12}\ M_\odot$ at $z=0$. The stellar distribution is modeled using a Hernquist profile, whose mass and half-mass radius were estimated by fitting the stellar masses and half-mass radii as a function of halo mass at $z=0$ based on our TNG50 MW sample. The upper left panel of Figure~\ref{fig09} shows the density profiles of FDM halos for different values of the particle mass $m_{22}$. As the boson mass decreases, the profiles become increasingly cored, with lower central densities and a flat core extending over several kpc. For $m_{22} < 1$, the central density is up to 3–4 orders of magnitude lower than that of the cuspy CDM profile (see Figure~\ref{fig09}).  
In contrast, for $m_{22} > 5$, FDM halos become denser than the classic NFW profile. A gradual convergence toward the NFW shape is also observed outside the central region. The $E$-$L_{z}$ diagram for FDM, computed in the total gravitational potential (DM + stars) (see lower left panel of Figure~\ref{fig09}), displays the same behavior previously identified. For low $m_{22}$, the orbital phase space is reduced compared to CDM, which enhances GC trapping in the central regions. Conversely, for higher $m_{22}$, GCs can access a much broader range of orbital energies.

\subsection{Self-interacting dark matter}

Another promising alternative is SIDM, proposed to address small-scale problems \citep{2000PhRvL..84.3760S,1992ApJ...398...43C,1995ApJ...452..495D}. In this scenario, DM particles are assumed to undergo isotropic elastic scatterings with a velocity-independent interaction cross-section. Self-interactions are usually quantified in terms of the cross-section per unit particle mass, $\sigma_m$, in units of cm$^2$/g. A generic prediction of SIDM is the formation of central DM cores whose size depends on the value of $\sigma_m$ \citep{2016PhRvL.116d1302K}. In the upper middle panel of Figure~\ref{fig09}, we show the SIDM halo profiles computed using the analytical model\footnote{Available at \url{https://github.com/JiangFangzhou/SIDM}} from \cite{Jiang23}. In our calculations, we assumed a constant, velocity-independent cross-section. Figure~\ref{fig09} reveals that the DM core size is inversely proportional to $\sigma_m$. SIDM produces larger and less dense cores across the range of model parameters than FDM. In addition, the $E$–$L_z$ diagram clearly shows that the sole dynamical impact of the DM core—combined with the baryonic component is a widening of the central valley as $\sigma_m$ increases (see bottom middle panel of Figure~\ref{fig09}). In contrast to FDM, SIDM does not confine GCs but rather allows them to occupy more extended orbits.

\subsection{Warm dark matter}

\begin{figure*}
    \centering
    \includegraphics[width=\linewidth]{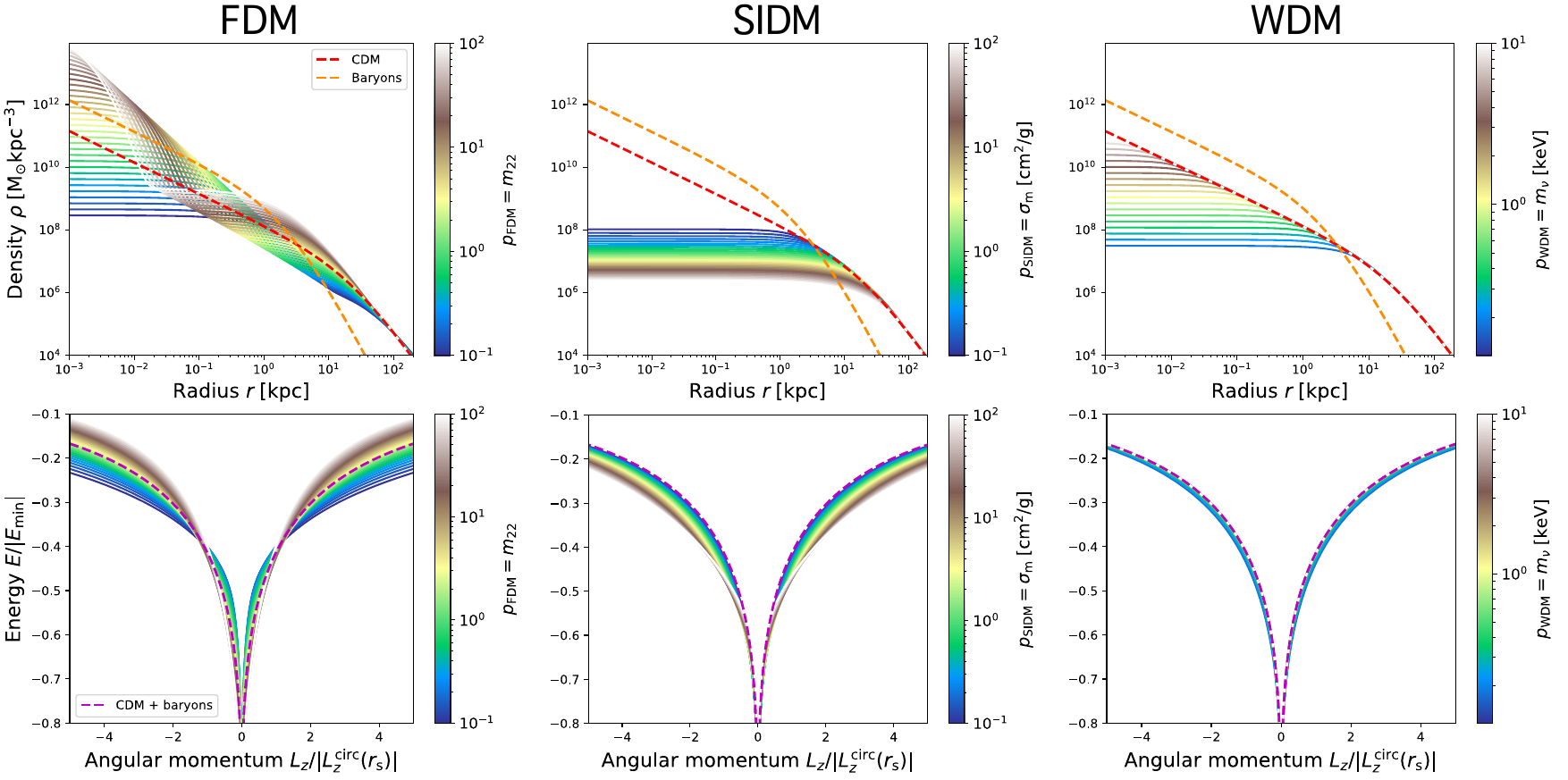}
    \caption{Comparison of the effects of different DM models (FDM, SIDM, WDM) on the structure of DM halos and on the orbital phase space accessible to in-situ GCs in an MW-like potential with $M_{\mathrm{halo}} = 10^{12}\ M_\odot$. Top row: DM density profiles for different parameter values of each model ($m_{22}$ for FDM, $\sigma/m$ for SIDM, and $m_\nu$ for WDM), in the absence of baryons. The red and orange dashed lines represent the CDM and baryonic profiles, respectively. Bottom row: Normalized $E$–$L_z$ diagrams computed in the total gravitational potential (DM + stars) using the same parameters as in the top row. These diagrams describe the orbital phase space accessible to in-situ GCs at $z = 0$, which varies depending on the DM model parameter.}
    \label{fig09}
\end{figure*}

Sterile neutrinos with masses in the keV range are a promising WDM candidate. Unlike CDM, WDM particles have a non-negligible thermal velocity. These velocities are large enough for particles to escape from small initial overdensities, thereby suppressing the formation of small-scale structures. This results in a cutoff in the matter power spectrum, where structures below a certain scale are erased \citep{1980PhRvL..45.1980B,1994PhRvL..72...17D,2000PhRvD..62f3511H,2006PhRvD..73f3506A,2005PhRvD..71f3534V}. Another important feature of this fermionic DM scenario is the natural emergence of cored matter density profiles \citep{2000ApJ...542..622C,2001ApJ...556...93B,2012MNRAS.424..684S,Maccio12,2013MNRAS.430.2346S,2014MNRAS.439..300L}. To model WDM density profiles, we used the CORENFW profile from \cite{2017MNRAS.467.2019R}, where the DM core size was computed using the analytical formula from \cite{Maccio12} (see Figure~\ref{fig09}). The profile parameters were calibrated to match the simulation results of \cite{Maccio12}. In contrast to the other two alternative DM models (FDM and SIDM), for high particle masses of the order of 10 keV, WDM profiles converge entirely to NFW at all radii. Since large particle masses yield cuspy profiles, WDM produces extremely small DM cores—comparable in size to the largest SIDM cores. As a result, WDM cores have higher central densities than SIDM and are similar to FDM, but never exceed the central density of an equivalent NFW profile. The $E$–$L_z$ diagram in Figure~\ref{fig09} shows very little deviation from CDM. The phase-space structure is nearly indistinguishable (see bottom right panel of Figure~\ref{fig09}). Indeed, in the presence of baryons, WDM has virtually no effect on the orbital dynamics of in-situ GCs in massive halos like those of MW analogs. Our results also showed that the orbital confinement phenomenon observed for GCs is absent in both SIDM and WDM, and thus appears to be a unique dynamical signature of FDM.

\end{document}